\documentclass[useAMS]{gGAF2e}
\usepackage{epsfig, journals}

\usepackage{amsmath}
\usepackage{amssymb}

\usepackage{multiequations}

\newcommand*{\be}{\begin{equation}}
\newcommand*{\ee}{\end{equation}}
\newcommand*{\bea}{\begin{eqnarray}}
\newcommand*{\eea}{\end{eqnarray}}
\newcommand*{\bme}{\begin{multiequations}}
\newcommand*{\eme}{\end{multiequations}}
\newcommand*{\se}{\singleequation}

\newcommand*{\qe}{\quadequation}

\newcommand*{\eR}{{\mathrm e}}
\newcommand*{\iR}{{\mathrm i}}

\newcommand{\citen}[1]{{\citealt{#1}}}

\newcommand{\vc}[1]{\mbox{\bf #1}}
\newcommand{\pdv}[2]{\frac{\partial #1}{\partial #2}}
\newcommand{\ppd}[2]{\frac{\partial^2 #1}{\partial {#2}^2}}

\newcommand{\NN}{\mathcal{N}}
\newcommand{\tbeta}{{\tilde\sigma}}
\newcommand{\tS}{{\tilde s}}
\newcommand{\tQ}{{\tilde q}}
\newcommand{\deltar}{\delta_R}
\newcommand{\deltai}{\delta_I}

\begin{document}
\doi{10.1080/03091920xxxxxxxxx}
 \issn{1029-0419} \issnp{0309-1929} \jvol{00} \jnum{00} \jyear{2009} 

\markboth{Petrovay, Kerekes \& Erd\'elyi}{Interface dynamo over finite shear layer}

\title{An analytic interface dynamo over a shear layer of finite depth}
\author{K. Petrovay${\dag}$$^{\ast}$\thanks{$^\ast$Corresponding author. E-mail: 
K.Petrovay@astro.elte.hu}, 
A. Kerekes${\ddag}$ and R. Erd{\'el}yi${\ddag}$\\
\vspace{6pt}  
${\dag}$E\"otv\"os University, Department of Astronomy, Budapest, Pf. 32, 
    H-1518 Hungary, and\\
ASIAA/National Tsing Hua University - TIARA, Hsinchu, Taiwan\\
$\ddag$SP$^2$RC, Dept. of Applied Mathematics, Univ. of Sheffield, 
Hicks Building, Hounsfield Road, Sheffield S3 7RH, UK
\\\vspace{6pt}\received{} 
}

\maketitle

\begin{abstract}
Parker's analytic Cartesian interface dynamo is generalized to the case of a
shear layer of finite thickness and low resistivity (``tachocline''), bounded
by a perfect conductor (``radiative zone'') on the one side, and by a highly
diffusive medium (``convective zone'') supporting an $\alpha$-effect on
the other side. In the limit of high diffusivity contrast between the shear
layer and the diffusive medium, thought to be relevant for the Sun, a pair of
exact dispersion relations for the growth rate and frequency of dynamo modes is
analytically derived. Graphic solution of the dispersion relations displays a
somewhat unexpected, non-monotonic behaviour, the mathematical origin of which
is elucidated. The dependence of the results on the parameter values (dynamo
number and shear layer thickness) is investigated.  The implications of this
result for the solar dynamo problem are discussed.
\end{abstract}


\section{Introduction}

Interface dynamos are a widely discussed class of astrophysical dynamos,
especially in the solar context (\citen{Petrovay:SOLSPA}, \citen{Solanki+:RPP},
\citen{Charbonneau:livingreview}). In contrast to the textbook case of Parker's
migratory dynamo, in an interface dynamo the $\alpha$ and $\Omega$ effects
operate in spatially distinct but adjacent layers. In this setup the dynamo wave
will arise as a surface wave propagating along the interface of the two regions.
While dynamo problems with spatially separated $\alpha$ and $\Omega$ had been
considered since the 1970's (see e.g. the references in
\citen{Zeldovich+:mg.book}), the classic analytic formulation of the problem is
due to \cite{Parker:interface}. 

It has long been suspected (\citen{Parker:buoy.prob},
\citen{Schussler:vort.pump}, \citen{Petrovay:Helsinki}), and recently
demonstrated in numerical simulations (\citen{Browning+:dynsimu.pumping}) that
the toroidal magnetic flux generated in the solar dynamo can only be stored
below the base of the convective zone proper, as various flux transport effects
(buoyancy and pumping) remove it quite effectively from the convective zone.
With the discovery of the tachocline layer it became clear that this
hypothetical flux reservoir coincides with the strongest rotational shear in the
solar interior, which is plausibly also responsible for the production of the
strong toroidal magnetic fields, i.e.~that this is the site of the
$\Omega$-effect. The site and nature of the $\alpha$-effect is much less clear;
in interface dynamo models it is assumed to be concentrated in the deep
convective zone, just above the tachocline.

Current helioseismic evidence indicates that at low heliographic latitudes the 
tachocline is situated immediately below the adiabatically stratified convective
zone. (At higher latitudes some overlap may be present.) This suggests that the
tachocline is much less turbulent than the convective zone proper. This is also
consistent with another physical consideration: the high overall toroidal field
strength that must be present if all the magnetic flux emerging during a solar
cycle resides in the thin tachocline layer should strongly suppress turbulence
there. The strong subadiabatic stratification of the tachocline will then
inhibit the penetration of meridional circulation to it, presenting a difficulty
for flux transport models, the main contender of interface dynamo models.

There is some disagreement over whether the oscillatory magnetic field generated
by the dynamo pervades the whole of the tachocline or not. For effective
magnetic diffusivities less than $10^8\,$cm$^2/$s or so, the penetration depth
(skin depth) of the oscillatory magnetic field is much less than the
helioseismically inferred tachocline thickness (order of 10 Mm), so the
penetration cannot be complete. In this ``slow tachocline'' scenario
(\citen{Garaud:tacho1}, \citen{Brun+Zahn:slow.tacho}) the dynamical changes in
the bulk of the tachocline take place on the long diffusive timescale. In the
alternative ``fast tachocline'' scenario (\citen{Dajka+Petrovay:mgconf},
\citeyear{Dajka+Petrovay:fast1}, \citen{Dajka:fast2}), hydrodynamical
instabilities maintain some modest level of turbulence in the tachocline, so the
resulting higher skin depth allows the dynamo field to penetrate deeper. In this
case it is the Lorentz force in the dynamo generated field that limits the
penetration of differential rotation deeper into the radiative interior, so
tachocline thickness and skin depth are intimately related and should agree
within an order of unity factor (\citen{Petrovay:1D.tacho}). In the present
paper we will consider the fast tachocline case; some observational support for
this scenario comes from the recent detection of solar cycle related changes in
tachocline properties (\citen{Baldner+Basu:11yr.tacho}), as predicted by the
fast tachocline models.

The interface dynamo option for the solar dynamo is attractive, as it would
explain the high ($\sim 100\,$kG) field strength of toroidal fields and their
equatorward propagation, while still being consistent with physical assumptions
about the magnetic diffusivity values in both the convective zone and tachocline
and about the meridional flow amplitude.  Interface dynamo models for the Sun
have been constructed, e.g.,\ by \cite{Charbonneau+McGregor:IFdynamo},
\cite{Tobias:nonlin.interface} or \cite{Markiel+Thomas}. Unfortunately, none of
these models show a compelling detailed agreement with observed features of the
cycle. Their main alternative, the flux transport dynamos
(\citen{Dikpati+Char:BLdynamo}, \citen{Chatterjee+:parity}), in contrast, are
more straightforward to parameterize so that they can reproduce some salient
features of the solar cycle reasonably well; however, the physical consistency
of the models is very much in doubt, especially regarding their assumptions on
diffusive and advective magnetic flux transport processes
(\citen{Choudhuri:BLdynamo.problems}).

The now classic analytic \cite{Parker:interface} interface dynamo consisted of
two adjacent semiinfinite domains: one highly diffusive domain home to the
$\alpha$-effect and another, low diffusivity domain characterized by strong
shear. As we have seen, the solar tachocline, corresponding to the shear layer
in the Cartesian interface dynamo, has a quite limited thickness, so it is not
particularly well represented by a semiinfinite domain. The aim of the present
paper is to generalize Parker's analytical interface dynamo to the case when the
shear layer has a finite thickness.

What makes this problem particularly relevant is the recent proposal
(\citen{Petrovay:bimodal}) that combining an interface model with a fast
tachocline scenario both the magnetic field amplitude and the thickness of the
tachocline can be determined (in the simplest case, by solving two algebraic
equations). Under some plausible assumptions, two solutions are found: one with
strong field and thin tachocline and another one with weaker field and a thicker
tachocline.  These represent ``high'' and ``low'' states of solar activity,
suggesting a link with the phenomenon of grand minima. Based on an analogy with
surface gravity waves, this model assumed that a finite shear layer thickness
$h$ reduces the growth rate of the interface dynamo by a factor $\tanh(hk)$,
where $k$ is the wave number. One objective of the present work is to examine
the validity of this assumption.

We note that interface dynamos with finite shear layers have previously been
considered in many numerical implementations
(e.g.~\citen{Charbonneau+McGregor:IFdynamo}, \citen{Markiel+Thomas}) as well as
in a semianalytic Cartesian setup more general than our current model
(\citen{Zhang+:sandwich.dynamo}). The effect of varying $h$, however, was not
studied in any of those models, due to the large number of other free parameters
and to the different focus of those papers. It is therefore legitimate to
consider a setup where $h$ is a variable parameter while in other respects the
model is kept as simple as possible.

The structure of our paper is as follows. In section 2 we analytically derive
the dispersion relations in the limit relevant to the Sun. Section 3 presents
graphical solutions of these dispersion relations and discusses their
properties. Finally section 4 concludes the paper.

\section{Derivation of the Dispersion Relations}

\subsection{Generic complex dispersion relations}

We use a Cartesian setup where the coordinate $z$ corresponds to height above
the bottom of the convective zone in the solar application, while $x$ and $y$
correspond to heliographic latitude and longitude, respectively. Let $\eta$
denote the magnetic diffusivity, $\alpha$ the $\alpha$ parameter of dynamo
theory (see e.g. \citen{Petrovay:SOLSPA}) and $\Omega=dv_y/dz$ the shear rate of
the velocity field representing radial differential rotation.  The magnetic
field is decomposed into a ``toroidal'' component $B_y$ and a
``poloidal'' component $\nabla\times A_y{\vc e}_y$ where $A_y$ is the
toroidal vector potential.

Consider now the following 3-layer ``sandwich'' setup:

\medskip

\begin{tabular}{lclcccc}
Convective zone & $z>0$ & $\eta=\eta_+$ & $\alpha\neq 0$ & $\Omega=0$ & 
  $B_y=B(x,z,t)$ & $A_y=A(x,z,t)$ \\
&&\\
Tachocline & $-h<z<0$ &  $\eta=\eta_-$ & $\alpha=0$ & $\Omega\neq 0$ & 
  $B_y=b(x,z,t)$ & $A_y=a(x,z,t)$ \\
&&\\
Radiative interior & $z<-h$ & $\eta=0$ & $\alpha=0$ & $\Omega=0$ & 
  $B_y=0$ & $A_y=0$ 
\end{tabular}
\medskip\\
The parameters $\eta_+$, $\eta_-$, $\alpha$ and $\Omega$ are assumed to be
constant in their respective layers.

The $\alpha\Omega$ dynamo equations in $z>0$ then read
\bme
\se
\label{eq:dynamo1}
\begin{eqnarray}
&&\pdv Bt - \eta_+\left(\ppd Bx+\ppd Bz\right)=0, \\
&&\pdv At - \eta_+\left(\ppd Ax+\ppd Az\right)=\alpha B , 
\end{eqnarray}
\eme
while in the tachocline we have
\bme
\se
\label{eq:dynamo2}
\begin{eqnarray}
&&\pdv bt - \eta_-\left(\ppd bx+\ppd bz\right)=\Omega\pdv ax, \\
&&\pdv at - \eta_-\left(\ppd ax+\ppd az\right)=0 . 
\end{eqnarray}
\eme

The matching conditions at $z=0$ are
\bme
\label{eq:fitcond1}
\begin{equation}
\qe
b=B, \qquad\quad a=A, \qquad\quad  \pdv az=\pdv Az, \qquad\quad  \eta_-\pdv bz = \eta_+\pdv Bz , 
\end{equation}
\eme
while at $z=-h$ they simply read
\bme
\label{eq:fitcond2}
\begin{equation}
b=0, \qquad\qquad\qquad a=0 .  
\end{equation}
\eme

Following the standard procedure for the analytical solution of interface dynamo
equations (\citen{Parker:interface}, \citen{Petrovay+Kerekes:intflow}), for the
solutions in terms of our variables $f(x,z,t)$ we consider normal modes of the
form $f(z)\exp[(\sigma+\iR\omega)t+\iR kx]$. In $z>0$, the $z$-dependent part $f(z)$ 
is sought in the form
\bme
\label{eq:ansatz1}
\begin{equation}
 B=C\eR^{-Rz}  \qquad\qquad\qquad A=(D+Ez)\eR^{-Rz} , 
\end{equation}
\eme
where $R=S+\iR Q$, and $C$, $\sigma$, $\omega$, $k$, $S$ and $Q$ are all real,
while $D$ and $E$ are complex. As we expect the field to vanish as
$z\rightarrow\infty$, $S$ must be positive.

In the shear layer the solutions have a similar form, but, owing to the finite
thickness of the layer, the solution is in general a superposition of modes
growing and decaying with $z$:
\bme
\label{eq:ansatz2}
\begin{equation}
 a=(J \eR^{rz}+J_1 \eR^{-rz}), \qquad\qquad  b=(L+Mz)\eR^{rz}+(L_1+M_1 z)\eR^{-rz} ,
\end{equation}
\eme
where $r=s+\iR q$ with $s$ and $q$ real and $s>0$, as before. Equations
(\ref{eq:fitcond2}a,b) then imply
\bme
\label{eq:fitcond2b}
\begin{equation}
 J_1=-J \eR^{-2rh}, \qquad\qquad  L_1=h M_1 +(hM-L)\eR^{-2hr} ,
\end{equation}
\eme
respectively.

Substitution of our trial solutions (\ref{eq:ansatz1}a,b), (\ref{eq:ansatz2}a,b) into the four interface fitting conditions 
(\ref{eq:fitcond1}a-d) yields
\bme
\label{eq:interf}
\se
\begin{align}
 J(1-\eR^{-2hr})&=D, \\
 L(1-\eR^{-2hr})+h(M \eR^{-2hr}+M_1)&=C ,\\ 
 rJ(1+\eR^{-2hr})&=E-DR ,\\
 \eta_-\left[ rL(1+\eR^{-2hr})+M+M_1-rh(M \eR^{-2hr}+M_1)\right] &= -\eta_+ RC .
\end{align}
\eme

On the other hand, substituting (\ref{eq:fitcond2b}a,b) into the trial solutions
(\ref{eq:ansatz1}a,b) and  (\ref{eq:ansatz2}a,b), and substituting these into the
dynamo equations  (\ref{eq:dynamo1}a,b), (\ref{eq:dynamo2}a,b) 
we obtain the dispersion
relations
\bme
\label{eq:dispr}
\se
\begin{align}
 \eta_+ R^2&=\sigma+\iR\omega+\eta_+k^2,\\
 \eta_- r^2&=\sigma+\iR\omega+\eta_-k^2,\\
  E&=\alpha C/2\eta_+ R,\\ 
 -2\eta_-r(M\eR^{rz}-M_1\eR^{-rz}) &= \iR k\Omega J(\eR^{rz}-\eR^{-rz}\eR^{-2hr})
\end{align}
\eme
the last of which implies the two separate relations
\bme
\begin{equation}
M=-\iR k\Omega J/2\eta_- r, \qquad\qquad\qquad M_1=M \eR^{-2hr} .
\end{equation}
\eme

Using the relations (\ref{eq:interf}a-c),
as well as (\ref{eq:dispr}c,d),
equation (\ref{eq:interf}d)
can be written as
\begin{equation}
rR(\mu^2 r+R\delta)(r+R\delta)/k^4 =\tfrac 14 \iR N [\delta-(1-\delta^2)hr] ,
  \label{eq:compdisp1}
\end{equation}
where we introduced the dynamo number $N=\alpha\Omega/\eta_+^2 k^3$ and the
notation $\delta=\tanh(hr)$. Equation (\ref{eq:compdisp1}) establishes a
relation between the complex vertical wavenumbers $r$ and $R$. A second such
relation is derived by eliminating $(\sigma+\iR\omega)$ from equations
(\ref{eq:dispr}a,b):
\begin{equation}
R^2=\mu^2r^2 - (\mu^2-1)k^2 ,  \label{eq:compdisp2}
\end{equation}
where the diffusivity contrast $\mu^2=\eta_-/\eta_+$ was introduced.

Given the parameters $N$, $\mu$ and $\delta$ (or, equivalently, $h$) defining
the problem, the pair of complex dispersion relations
(\ref{eq:compdisp1}), (\ref{eq:compdisp2}) can in principle be solved for $r$
and $R$. The frequency $\omega$ and growth rate $\sigma$ of the dynamo wave then
follow from (\ref{eq:dispr}a,b).
The nine parameters $C$,
$D$, $E$, $J$, $J_1$, $L$, $L_1$, $M$ and $M_1$ are then determined by the
relations  (\ref{eq:fitcond2b})--(\ref{eq:interf}c) and 
(\ref{eq:dispr}c,d).
Substituting these into
(\ref{eq:ansatz1}a,b), (\ref{eq:ansatz2}a,b) together with the derived values of $r$
and $R$ finally should yield the full $z$-dependence of the solution for a given
value of $k$.

\subsection{Solar limit}

Unfortunately, an attempt to actually implement this procedure analytically
encounters difficulties. Equations (\ref{eq:compdisp1}), (\ref{eq:compdisp2})
look deceivingly simple, yet the actual calculation soon becomes quite involved.
To make progress, from this point onwards we restrict attention to the limit
relevant for the solar interior. 

In the solar convective zone the dynamo wave number and frequency are
empirically known to be of order $k\sim 1/R_\odot$ and $\omega\sim
\eta_+/R_\odot^2\sim\eta_+ k^2$, where $R_\odot$ is the solar radius. In the
tachocline below, the magnetic diffusivity is uncertain but surely several
orders magnitude lower than the fixed value $\eta_+$: we may then safely take
the limit $\eta_-\rightarrow 0$.  With the fixed values of $\eta_+$, $k$ and
$\omega$ given above, the penetration depth (skin depth) of the oscillatory
dynamo magnetic field will be $(2\eta_-/\omega)^{1/2}\sim\mu/k$, which also
implies $|r|\sim k/\mu$. As the thickness $h$ of a fast tachocline is expected
to be a few skin depths, so in this limit we also expect $h\sim\mu/k\rightarrow
0$, suggesting the use of a fixed nondimensional layer depth $H=hk/\mu$ instead.
Finally, inspection of equation (\ref{eq:compdisp1}) shows that the leading
terms on the left-hand side will be of order $1/\mu^2$, while those on the right-hand side are of
order $N$. So in order to keep the dynamo supercritical in the solar limit we
need to keep $\mu^2 N$ fixed.

Summing up these considerations, the limit relevant to the solar case is
\begin{equation}
\mu\rightarrow 0, \qquad \mbox{ while } \qquad H=hk/\mu=\mbox{finite} \qquad
\mbox { and } \qquad \NN=\mu^2 N=\mbox{finite.}
\end{equation}
Equation (\ref{eq:compdisp2}) then simplifies to
\begin{equation}
 R=(\mu^2r^2+k^2)^{1/2}\,.
\end{equation}
Substituting this into (\ref{eq:compdisp1}) yields an equation for the complex
wavenumber $r$ only. In analogy with the nondimensional layer depth $H$, we now
introduce the nondimensional equivalent of $r$ as $\Psi=\mu r/k$: clearly,
$\Psi$ will remain finite in the solar limit, as discussed above. Expressing
$r$ with $\Psi$ and keeping only leading order terms in the limit of 
$\mu\rightarrow 0$, equation (\ref{eq:compdisp1}) finally takes the form
\begin{equation}
  \left[\Psi^2(\delta^2-1)^2 H^2 + 2\Psi\delta(\delta^2-1)H +\delta^2\right]
  \NN^2/16 + 
  \iR\Psi^2(\Psi^2-1)\delta\left[\Psi(\delta^2-1)H + \delta\right]\NN/2 =
  \delta^2\Psi^4(1+\Psi^2)^2 .    \label{eq:psieq}  
\end{equation}
In this equation, $\Psi$ and $\delta$ are complex, while the parameters $\NN$
and $H$ are real. We write $\delta$ and $\Psi$ explicitly as
\bme
\label{eq:Psibetanu}
\begin{equation}
\delta=\delta_R+i\delta_I , \qquad\qquad\qquad  \Psi^2=\tbeta+i\nu ,
\end{equation}
\eme
where
\bme
\begin{equation}
\tbeta=\sigma/\eta_+ k^2 ,  \qquad\qquad\qquad   \nu=\omega/\eta_+ k^2
\end{equation}
\eme
are the dimensionless growth rate and frequency of the dynamo wave, and (\ref{eq:Psibetanu}b)
follows from 
(\ref{eq:dispr}b) 
in the limit $\mu\rightarrow 0$.

Upon performing these substitutions, equation (\ref{eq:psieq}) is split into
real and imaginary parts. (The calculation is very tedious but still manageable
by numerical algebra packages.) This finally yields two dispersion
relations for the nondimensional growth rate $\tbeta=\beta-1$ and the
nondimensional frequency $\nu$. These are of the form
\bme
\label{eq:disprel}
\begin{equation}
f_1(\tbeta,\nu;H,\NN)=0, 
\qquad\qquad\qquad\qquad
f_2(\tbeta,\nu;H,\NN)=0 , 
\end{equation}
\eme
where
\bme
\begin{equation}
f_1(\tbeta,\nu;H,\NN) \equiv (a_{11} H^2-a_{12})\NN^2+b_1 \NN+c_1,\qquad
f_2(\tbeta,\nu;H,\NN) \equiv (a_{21} H^2-a_{22})\NN^2+b_2 \NN+c_2
\end{equation}
\eme
with
\bme
\se
\begin{eqnarray}
a_{11} &=& \nu(1+\deltar^4+\deltai^4)
           -4\tbeta\deltar\deltai(1-\deltar^2+\deltai^2)
	   -2\nu(\deltar^2+3\deltar^2\deltai^2-\deltai^2),
\\
a_{12} &=& 2\deltai\deltar,
\\
b_1 &=& 8[ 2\nu(1+2\tbeta)\deltar\deltai
           -(\tbeta^2+\tbeta-\nu^2)(\deltar^2-\deltai^2)],
\\
c_1 &=& 32[(\tbeta^2+\tbeta-\nu^2)\deltar -\nu(2\tbeta+1)\deltai]
          [(\tbeta^2+\tbeta-\nu^2)\deltai +\nu(2\tbeta+1)\deltar],
\\
a_{21} &=& \tbeta(1+\deltar^4+\deltai^4)
           -4\nu\deltar\deltai(1-\deltar^2+\deltai^2)
	   -2\tbeta(\deltar^2+3\deltar^2\deltai^2-\deltai^2),
\\
a_{22} &=& \deltar^2-\deltai^2,
\\
b_2 &=& 8[ \nu(1+2\tbeta)(\deltar^2-\deltai^2)
           +2(\tbeta^2+\tbeta-\nu^2)\deltar\deltai],
\\
c_2&=&16[\tbeta(\tbeta+1)(\deltar-\deltai)-\nu(\nu+1)\deltar+\nu(\nu-1)\deltai]\nonumber
   \\
   &&  \times[\tbeta(\tbeta+1)(\deltar+\deltai)-\nu(\nu-1)\deltar+\nu(\nu+1)\deltai].
\end{eqnarray}
\eme

Recalling that $\delta=\tanh(hr)=\tanh(H\Psi)$, the real and imaginary parts of
$\delta$ are
\bme
\be
\deltar = \frac{\cosh(H\tS)\sinh(H\tS)}{\cosh^2(H\tS)-\sin^2(H\tQ)},
\qquad\qquad
\deltai = \frac{\cos(H\tQ)\sin(H\tQ)}{\cosh^2(H\tS)-\sin^2(H\tQ)} ,
\ee
\eme
where $\tS=\mu s/k$ and $\tQ=\mu q/k$ are related to $\tbeta$ and $\nu$ by
virtue of equations (\ref{eq:dispr}a,b):
\bme
\be
\tS^2 = \tfrac 12[\tbeta+(\tbeta^2+\nu^2)^{1/2}],    
\qquad\qquad\qquad
\tQ^2 = \frac{\nu^2}{\tbeta+[\tbeta^2+\nu^2]^{1/2}} .
\ee
\eme

It is straightforward to check that, in the limit $H\rightarrow\infty$ (i.e.
$\delta\rightarrow 1$), Parker's original solution is retrieved (cf.
equations (60), (61) in \citen{Parker:interface}).

\section{Solution}

While the equations (\ref{eq:disprel}a,b) are quadratic in $\NN$, they are in
general extremely complicated functions of the other variables. Nevertheless,
equations (\ref{eq:disprel}a,b) lend themselves most readily to a graphic
solution.  Fixing the value of the control parameters $\NN$ and $H$, solutions
correspond to the crossing points of the zero contours of functions $f_1$ and
$f_2$ in the $\nu\tbeta$ plane. Then, varying $H$ while keeping either $\NN$ or
the total shear fixed one can study how the solutions deviate from the Parker
solution ($H\rightarrow\infty$ limit) for finite values of $H$.

\begin{figure}
\centerline{
\epsfig{figure=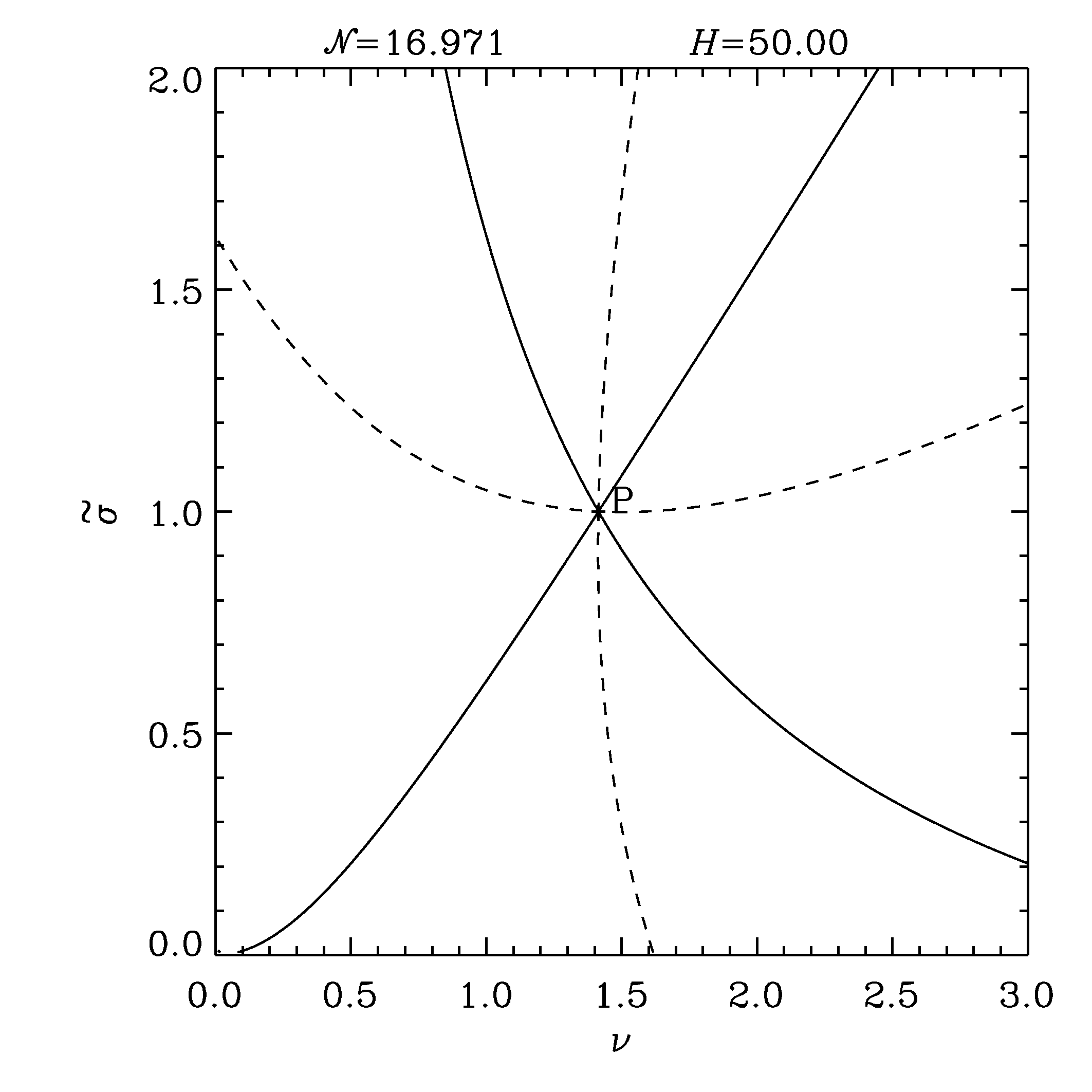,width=84mm,height=75mm}\hfill
\epsfig{figure=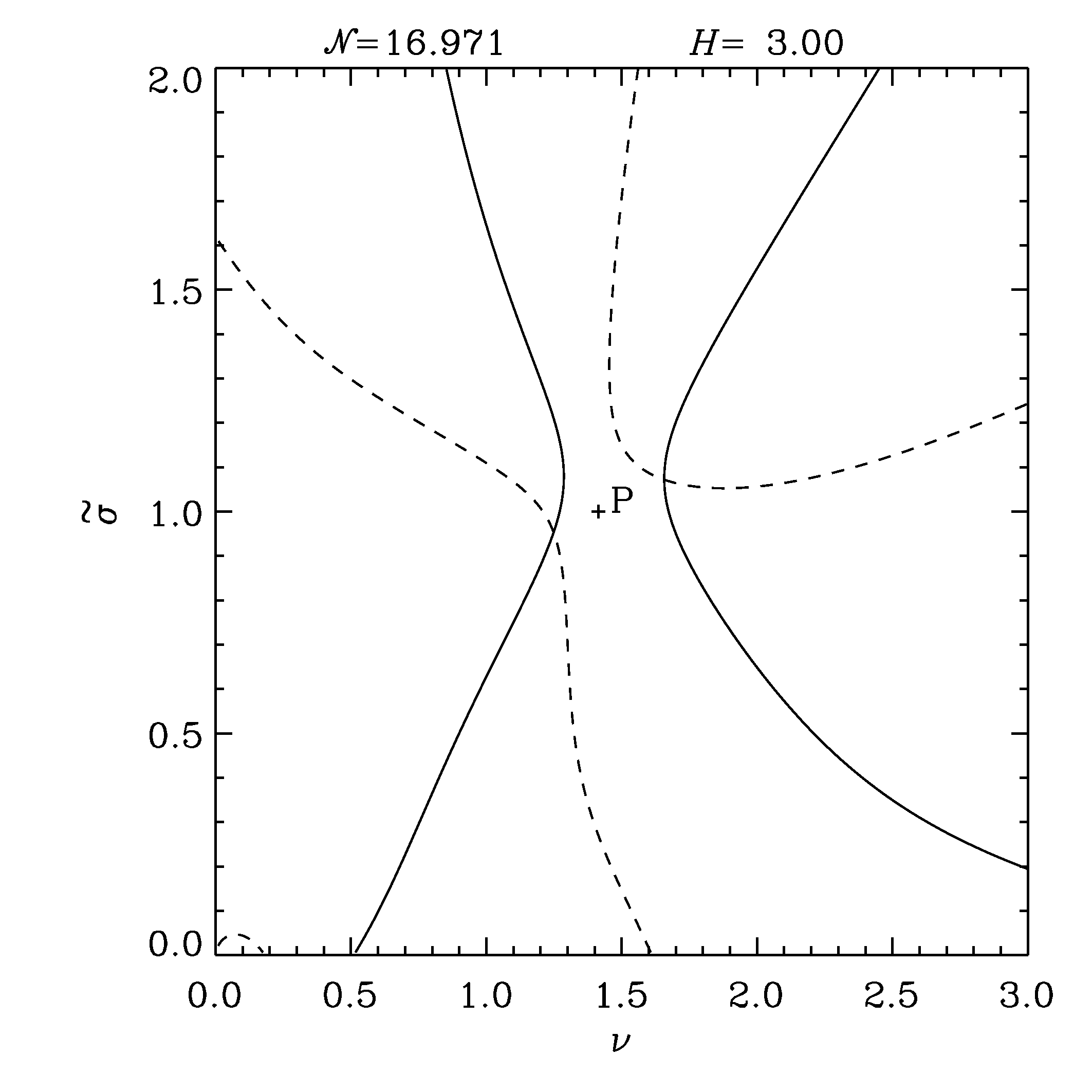,width=84mm,height=75mm}
}
\centerline{
\epsfig{figure=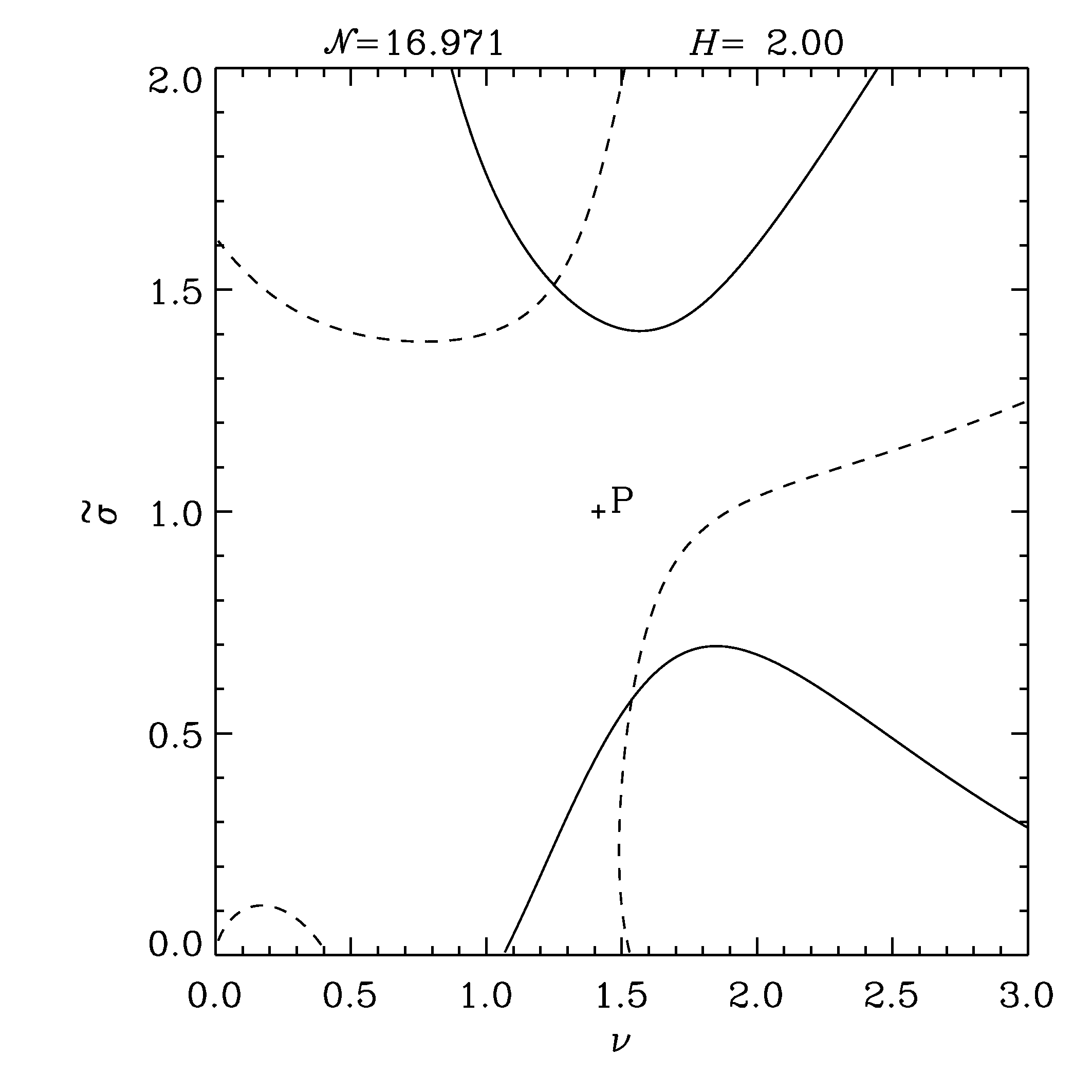,width=84mm,height=75mm}\hfill
\epsfig{figure=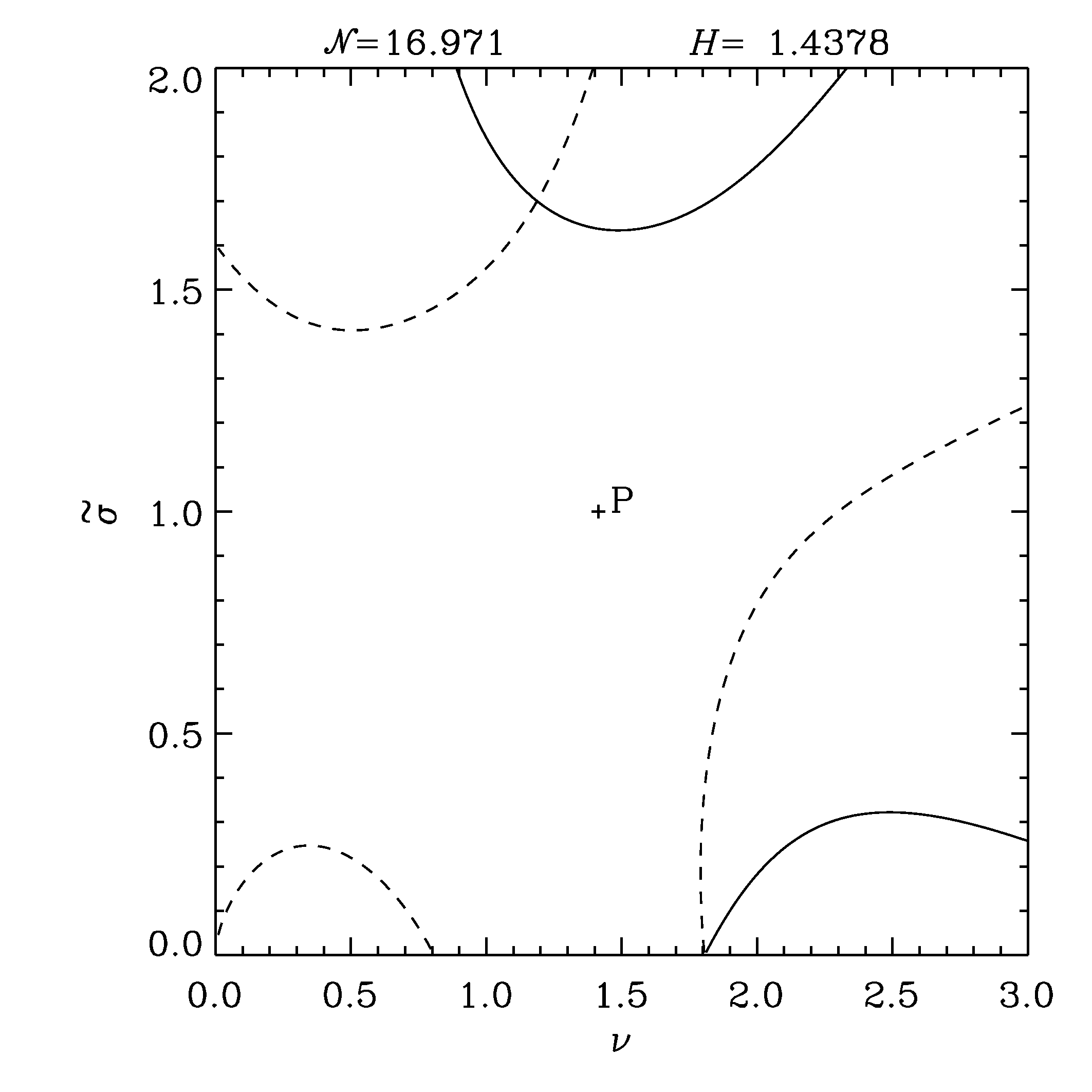,width=84mm,height=75mm}
}
\centerline{
\epsfig{figure=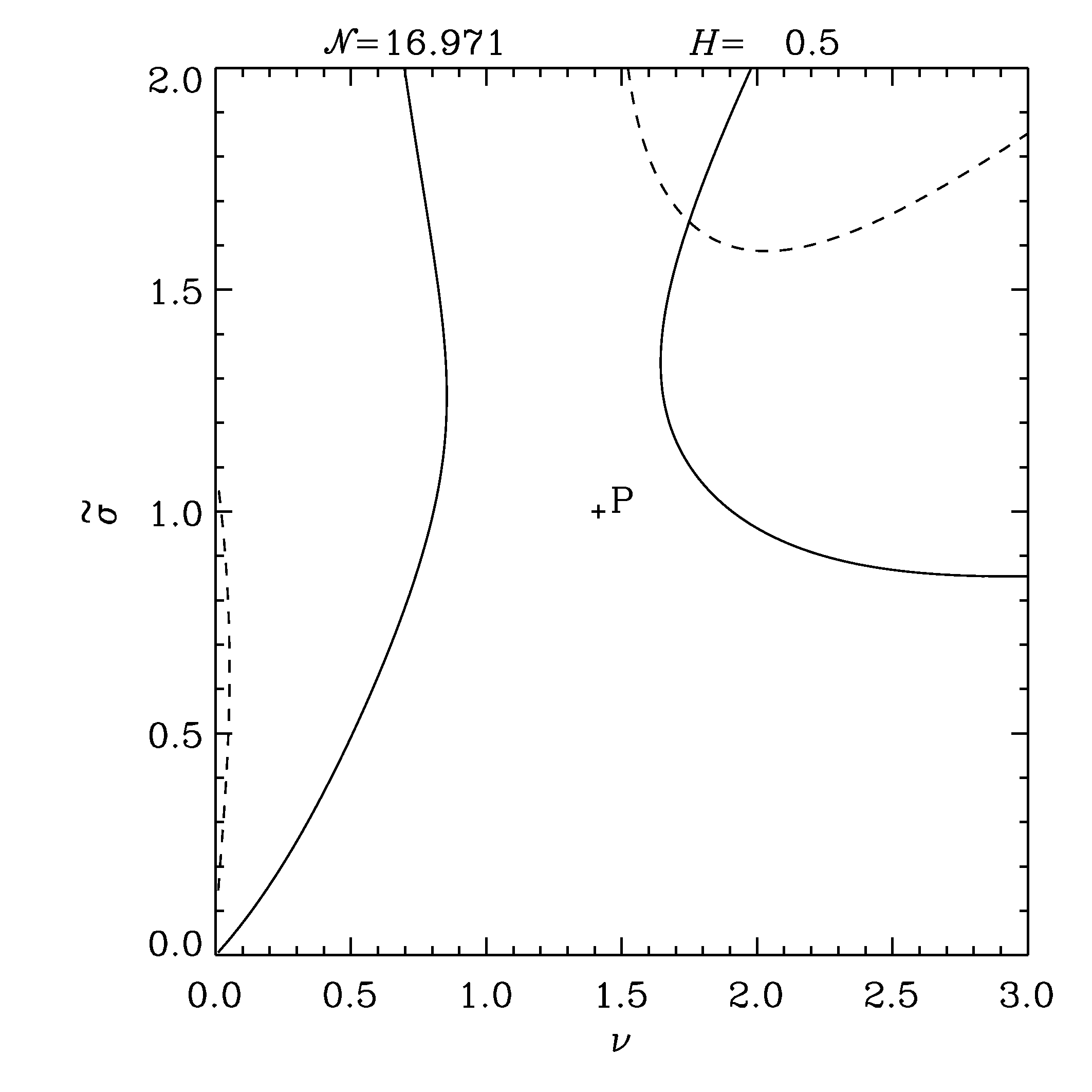,width=84mm,height=75mm}\hfill
\epsfig{figure=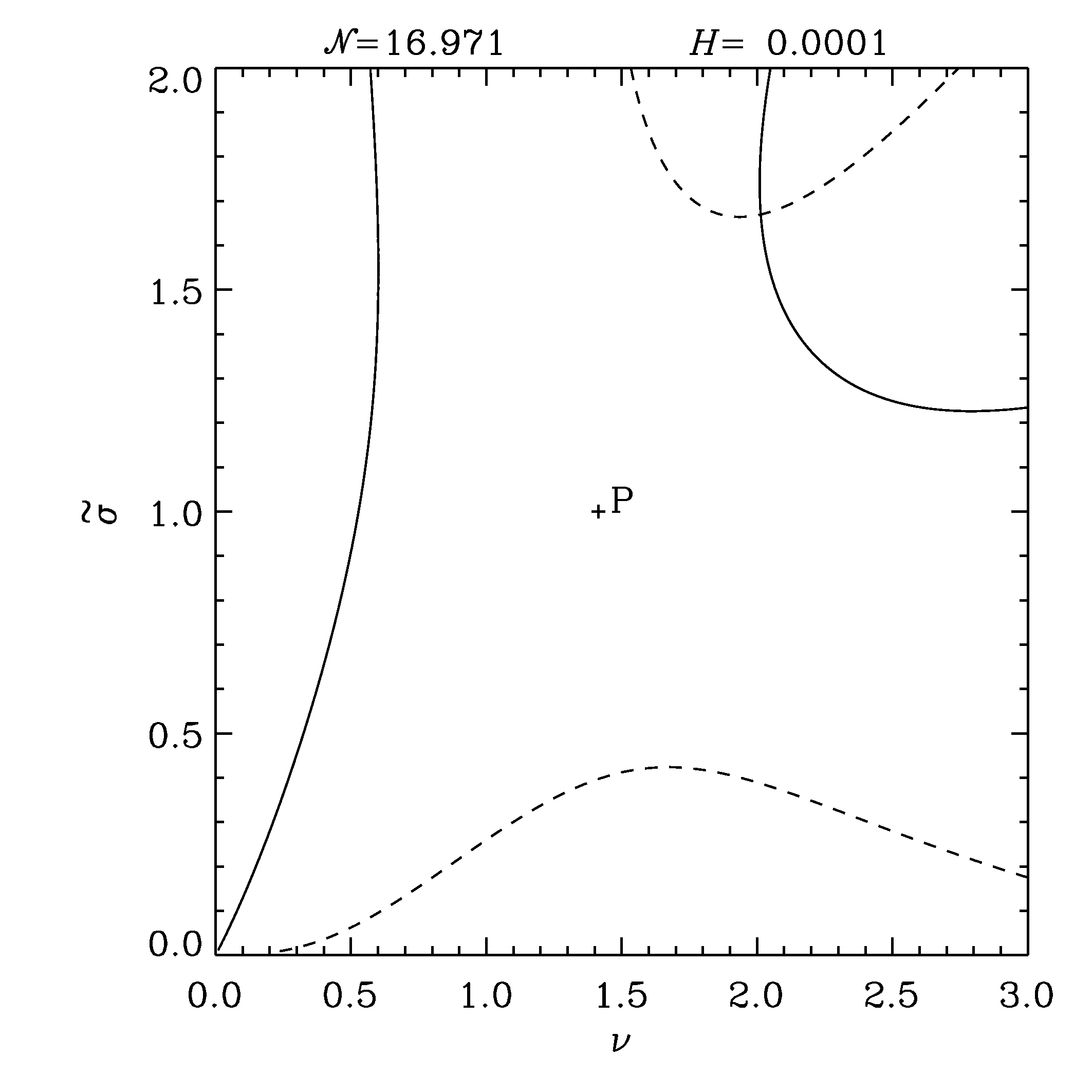,width=84mm,height=75mm}
}
\caption{Zero contours of the functions $f_1$ (solid) and $f_2$ (dashed) on the
$\nu\tbeta$ plane, for different values of the layer depth $H$. Crossing points
of the solid and dashed curves represent formal solutions of the dispersion
relations. The solution in the limit $H\rightarrow\infty$ is marked by $P$.}
\label{fig:textbook}
\end{figure}

For the purposes of this study we take the ``textbook'' case $\NN=12\sqrt 2$,
already considered by \cite{Parker:interface}. In the limit
$H\rightarrow\infty$, considered by Parker, there is only one growing mode
characterized by $\tbeta=1$ and $\nu=\sqrt 2$, as borne out in
figure~\ref{fig:textbook}$a$. As $H$ is decreased, the solution bifurcates, and
two sequences of apparently growing modes appear. Recall, however, that during
the derivation of the dispersion relations in section 2, several steps may have
resulted in the appearance of spurious solutions. Indeed, on physical
grounds we expect to return Parker's solution in the $H\rightarrow\infty$ limit
but also to behave as $\tbeta\rightarrow 0$ at some finite positive value of
$H$. On the basis of  this consideration we must reject the ``upper'' (i.e.\
higher growth rate) sequence of solutions in figure~\ref{fig:textbook}, converging
to $\tbeta=1.67$, $\nu=2.01$, as spurious. The lower sequence, representing
physical growing modes, in turn disappears for $H<1.45$, i.e. the dynamo is
suppressed for such low layer depths. 

\begin{figure}
\centerline{
\epsfig{figure=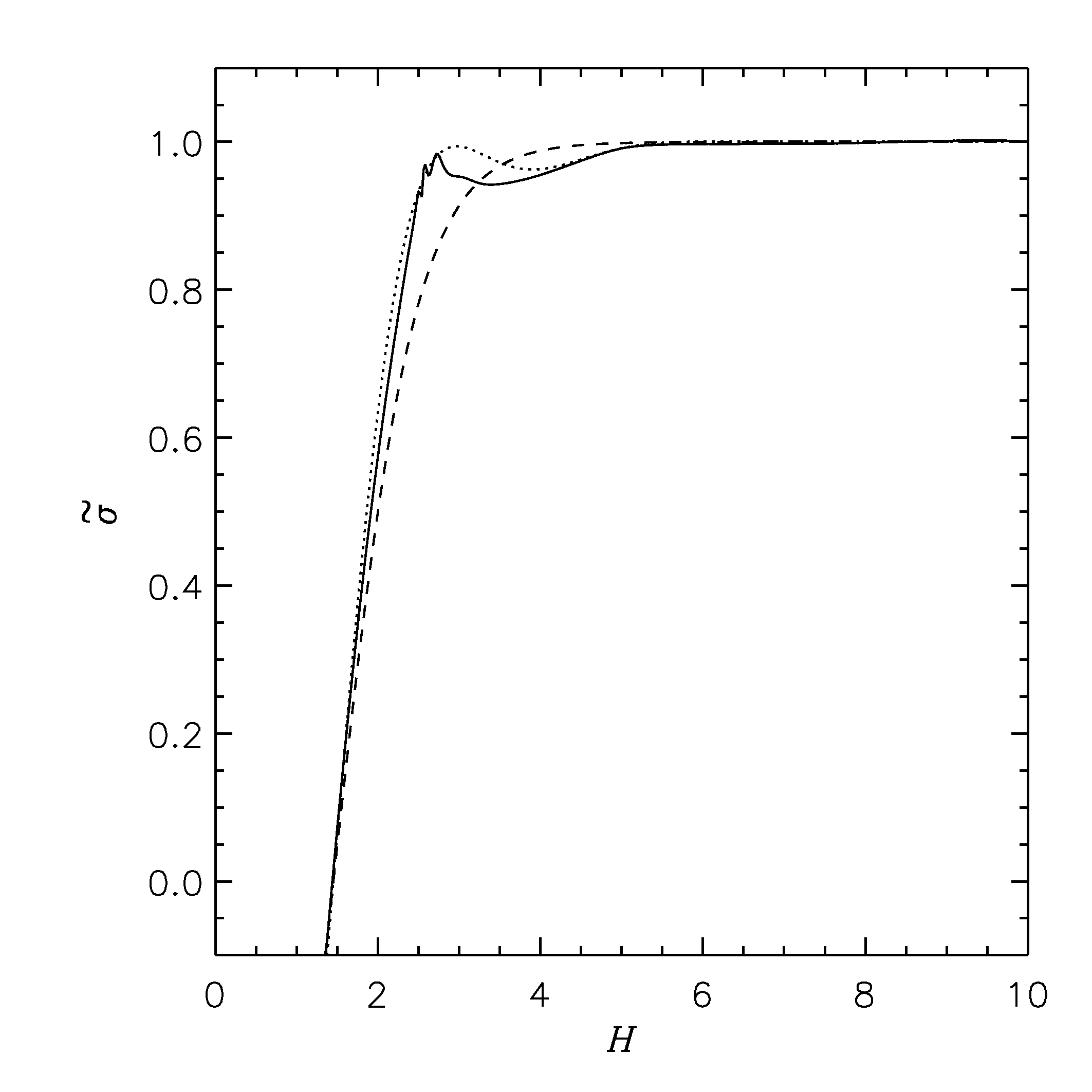,width=84mm}
}
\caption{Variation of the nondimensional growth rate $\tbeta$ with layer depth
$H$ for $\NN=12\sqrt 2$. The solution is analytically fitted by a
simple shifted tangent hyperbolic (dashed), on which a Gaussian damped
sinusoidal is superposed (dotted).}
\label{fig:textbook_curve}
\end{figure}

The variation of $\tbeta$ with $H$ is shown in figure~\ref{fig:textbook_curve}.
The variation can be roughly represented by a shifted tangent hyperbolic
function. This was to be expected in analogy with surface gravity waves
(\citen{Petrovay:bimodal}), the shift here being due to the suppression of
growing modes at a final layer depth.  The detailed behaviour of the solution
is, however, considerably richer that this simple theoretical expectation,
displaying a surprising non-monotonic modulation. This is a real physical
interference effect, due to the interaction of the waves originating at the
interface and the waves reflected from the bottom of the tachocline. To imitate
this modulation, we superpose an exponentially damped sinusoidal on the tangent
hyperbolic to arrive at the following fitting function
\bme
\begin{equation}
\tbeta=\tanh(H')+0.2\sin(1.5 H') \exp(-0.25 H'^2), \qquad\quad\mbox{where}\qquad H'=H-1.45 .
\end{equation}
\eme
Note, however, that the appearence and details of such interference effects may
be sensitive to the details of the model setup (i.e.\ to how sharp the bottom of
the tachocline is supposed to be).

\begin{figure}
\centerline{
\epsfig{figure=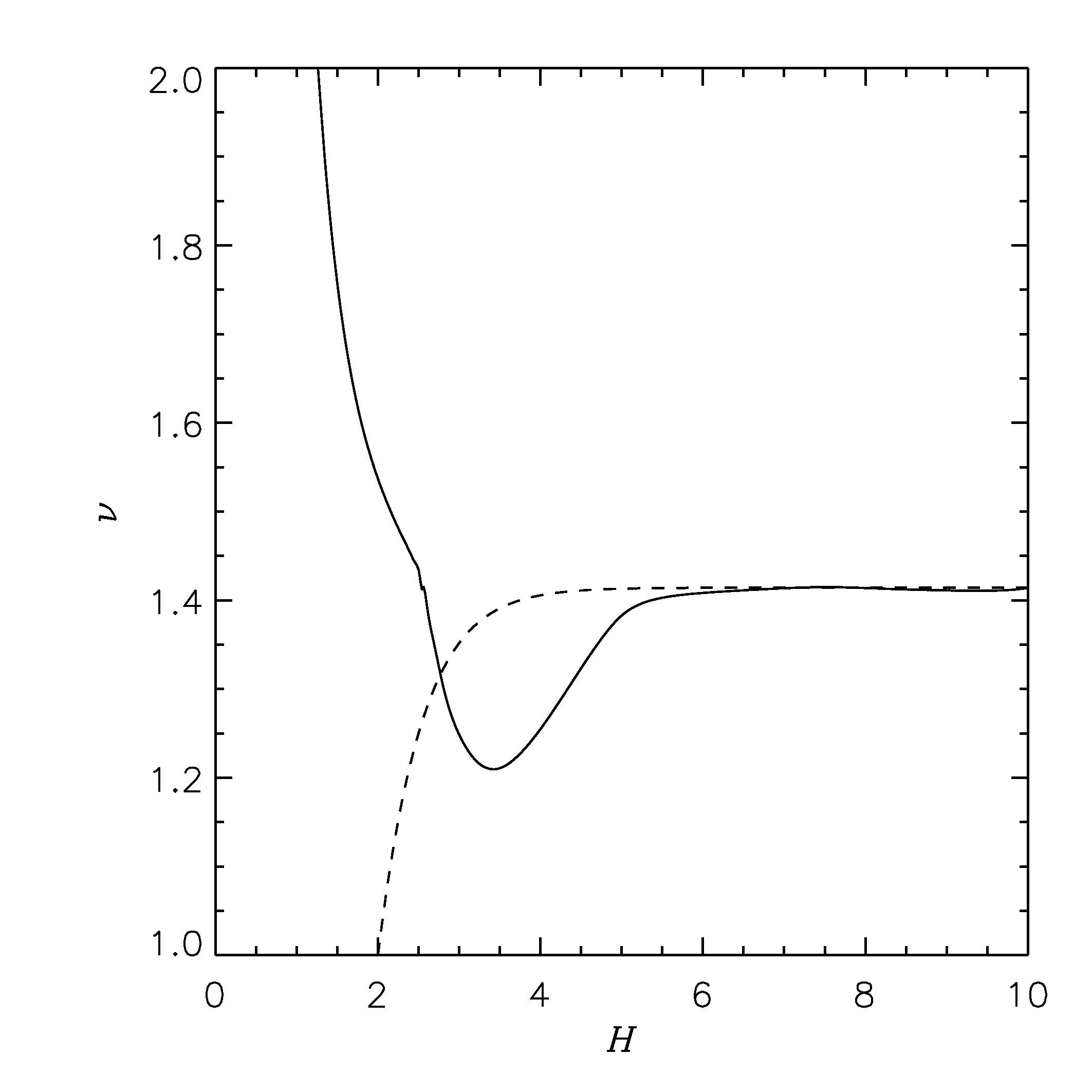,width=84mm}
}
\caption{Variation of the nondimensional frequency $\nu$ with layer depth
$H$ for $\NN=12\sqrt 2$. The relation $\nu^2\propto\tanh(H)$ (dashed), based on
an analogy with surface gravity waves, is shown for comparison.}
\label{fig:textbook_curve2}
\end{figure}

The breakdown of the water wave analogy is even more apparent from
figure~\ref{fig:textbook_curve2}. The variation of the frequency with layer depth
is clearly quite different for interface dynamo waves than the simple
$\nu^2\propto\tanh(H)$ relation (dashed), valid for surface gravity waves.

\begin{figure}
\centerline{
\epsfig{figure=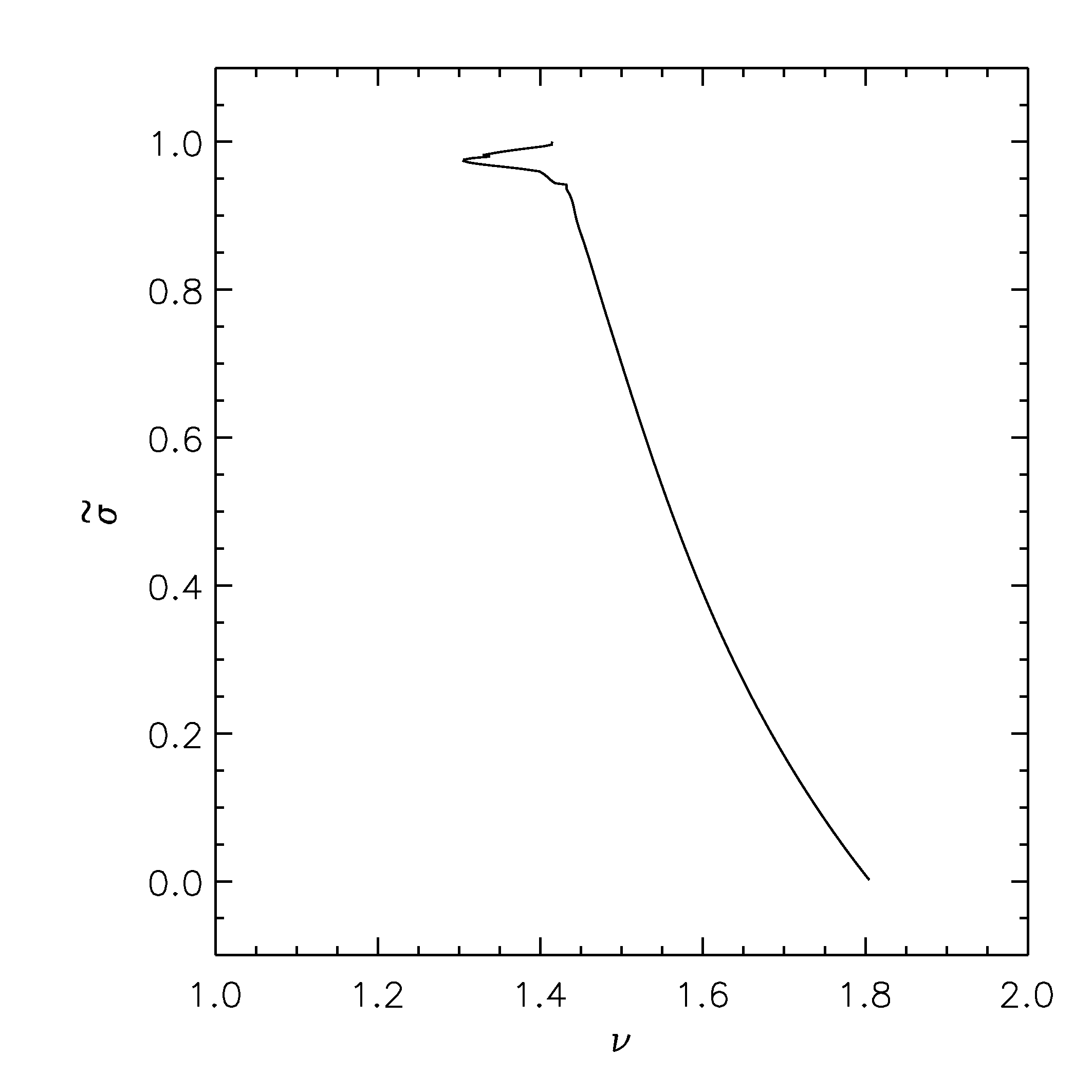,width=84mm}
}
\caption{Trajectory of the physical solution sequence in the $\nu\tbeta$ plane
for varying $H$ when $\NN=12\sqrt 2$ is kept fixed}
\label{fig:textbook_nubeta}
\end{figure}

Figure~\ref{fig:textbook_nubeta} presents the trajectory of the solution in the
$\nu$$\beta$ plane. It is apparent that in the critical case $\nu\simeq 1.8$, at
least an order of magnitude higher than the value $(2/N)^{1/2}$ suggested by
Parker's original analysis. 

\section{Conclusion}

In this paper we have generalized the textbook interface dynamo model of
\cite{Parker:interface} to the case of finite tachocline thickness, while
otherwise keeping the model as simple as possible. We found that the growth rate
of the dynamo has a shifted tangent hyperbolic dependence on the layer depth, in
agreement with the expectation based on an analogy with surface gravity waves.
This finding lends some support to the claim (\citen{Petrovay:bimodal}) that the
combination of a fast tachocline and an interface dynamo can give rise to
bimodal solutions, possibly explaining the phenomenon of grand minima. Under the
assumption that the growth rate of the dynamo has a dependence on the tachocline
depth similar to what was found in the present paper, in that previous work two
solutions were found for a nonlinear interface dynamo combined with a fast
tachocline: a strong field, thin tachocline solution corresponding to normal
solar activity and a weak field, thick tachocline solution that may potentially
represent a grand minimum state of solar acivity.

Yet the behaviour of the solutions does show some surprising features. The
nondimensional frequency  varies with layer depth in a non-intuitive way and the
detailed behaviour of the growth rate---layer depth curve displayes a
non-monotonic, oscillatory behaviour. This illustrates the richness of phenomena
appearing even in strongly simplified interface dynamo models. A further
systematic study of models of this type and of more generic models is needed to
more fully explore the parameter space and to obtain a deeper understanding of
the spectrum of possible dynamo solutions in an interface setup. Without such
systematic studies no final verdict on the viability of an interface-type
solution to the solar dynamo problem is possible.


\section*{Acknowledgments}
This research was supported by the Hungarian Science Research Fund (OTKA) under
grant no.\ K67746; by the European Commission through the SOLAIRE Network
(MTRN-CT-2006-035484); by the Theoretical Institute for Advanced Research in
Astrophysics (TIARA) operated under Academia Sinica and the National Science
Council Excellence Projects program in Taiwan administered through grant number
NSC95-2752-M-007-006-PAE; as well as by the Science and Technology Facilities
Council (STFC), UK and the Mathematics and Statistics Research Centre (MSRC) of
the University of Sheffield. K.P. acknowledges the warm hospitality received at
the Department of Applied Mathematics, University of Sheffield during his visit.
R.E. acknowledges M. K\'eray for patient encouragement.


\begin{thebibliography}{25}
\providecommand{\natexlab}[1]{#1}

\bibitem[\protect\citeauthoryear{{Baldner} and
  {Basu}}{2008}]{Baldner+Basu:11yr.tacho}
{Baldner}, C.S. and {Basu}, S., {Solar cycle related changes at the base of the
  convection zone}. {\itshape Astrophys.~J.}
2008, \textbf{686}, 1349--1361.

\bibitem[\protect\citeauthoryear{{Browning}
  {\itshape{et~al.}}}{2006}]{Browning+:dynsimu.pumping}
{Browning}, M.K., {Miesch}, M.S., {Brun}, A.S. and {Toomre}, J., {Dynamo action
  in the solar convection zone and tachocline: Pumping and organization of
  toroidal fields}. {\itshape Astrophys.~J.~Lett.}
2006, \textbf{648}, L157--L160.

\bibitem[\protect\citeauthoryear{{Brun} and
  {Zahn}}{2006}]{Brun+Zahn:slow.tacho}
{Brun}, A.S. and {Zahn}, J.P., {Magnetic confinement of the solar tachocline}.  {\itshape Astron.~Astrophys.}
2006, \textbf{457}, 665--674.

\bibitem[\protect\citeauthoryear{Charbonneau}{2005}]{Charbonneau:livingreview}
Charbonneau, P., {}. {\itshape Living Rev.\ Sol.\ Phys.} 2005, \textbf{2}, 2.

\bibitem[\protect\citeauthoryear{Charbonneau and
  MacGregor}{1997}]{Charbonneau+McGregor:IFdynamo}
Charbonneau, P. and MacGregor, K.B., {Solar interface dynamos II. Linear,
  kinematic models in spherical geometry}. {\itshape Astrophys.~J.}
1997, \textbf{486},
  502--520.

\bibitem[\protect\citeauthoryear{{Chatterjee}
  {\itshape{et~al.}}}{2004}]{Chatterjee+:parity}
{Chatterjee}, P., {Nandy}, D. and {Choudhuri}, A.R., {Full-sphere simulations
  of a circulation-dominated solar dynamo: Exploring the parity issue}. {\itshape Astron.~Astrophys.}
2004, \textbf{427}, 1019--1030.

\bibitem[\protect\citeauthoryear{{Choudhuri}}{2008}]{Choudhuri:BLdynamo.proble%
ms}
{Choudhuri}, A.R., {How far are we from a {``Standard Model''} of the solar
  dynamo?} {\itshape Adv.~Space Res.} 2008, \textbf{41}, 868--873.

\bibitem[\protect\citeauthoryear{{Dikpati} and
  {Charbonneau}}{1999}]{Dikpati+Char:BLdynamo}
{Dikpati}, M. and {Charbonneau}, P., {A Babcock-Leighton flux transport dynamo
  with solar-like differential rotation}. {\itshape Astrophys.~J.}
1999, \textbf{518},
  508--520.

\bibitem[\protect\citeauthoryear{{Forg\'acs-Dajka}}{2003}]{Dajka:fast2}
{Forg\'acs-Dajka}, E., {Dynamics of the fast solar tachocline II.} {\itshape Astron.~Astrophys.}
2003, \textbf{413}, 1143--1151.

\bibitem[\protect\citeauthoryear{{Forg\'acs-Dajka} and
  Petrovay}{2001}]{Dajka+Petrovay:mgconf}
{Forg\'acs-Dajka}, E. and Petrovay, K., {Tachocline confinement by an
  oscillatory magnetic field}. {\itshape \solphys} 2001, \textbf{203},
  195--210.

\bibitem[\protect\citeauthoryear{{Forg\'acs-Dajka} and
  Petrovay}{2002}]{Dajka+Petrovay:fast1}
{Forg\'acs-Dajka}, E. and Petrovay, K., {Dynamics of the fast solar tachocline
  I. Dipolar field}. {\itshape Astron.~Astrophys.}
2002, \textbf{389}, 629--640.

\bibitem[\protect\citeauthoryear{Garaud}{2001}]{Garaud:tacho1}
Garaud, P., {Dynamics of the solar tachocline I: An incompressible study}. {\itshape Mon.~Not.~Roy.~Astr.~Soc.}
2001, \textbf{329}, 1--18.

\bibitem[\protect\citeauthoryear{Markiel and Thomas}{1999}]{Markiel+Thomas}
Markiel, J.A. and Thomas, J.H., Solar interface dynamo models with a realistic
  rotation profile. {\itshape Astrophys.~J.}
1999, \textbf{523}, 827--837.

\bibitem[\protect\citeauthoryear{Parker}{1975}]{Parker:buoy.prob}
Parker, E.N., The generation of magnetic fields in astrophysical bodies X:
  Magnetic buoyancy and the solar dynamo. {\itshape Astrophys.~J.}
1975, \textbf{198},
  205--209.

\bibitem[\protect\citeauthoryear{{Parker}}{1993}]{Parker:interface}
{Parker}, E.N., {A solar dynamo surface wave at the interface between
  convection and nonuniform rotation}. {\itshape Astrophys.~J.}
1993, \textbf{408},
  707--719.

\bibitem[\protect\citeauthoryear{Petrovay}{1991}]{Petrovay:Helsinki}
Petrovay, K., Topological pumping in the lower overshoot layer; in {\itshape
  The Sun and Cool Stars: Activity, Magnetism, Dynamos}, edited by I.~Tuominen,
  D.~Moss and G.~R{\"u}diger 1991, pp. 67--70.

\bibitem[\protect\citeauthoryear{Petrovay}{2000}]{Petrovay:SOLSPA}
Petrovay, K., What makes the Sun tick?; in {\itshape The Solar Cycle and
  Terrestrial Climate} 2000, pp. 3--14.

\bibitem[\protect\citeauthoryear{{Petrovay}}{2003}]{Petrovay:1D.tacho}
{Petrovay}, K., {A consistent one-dimensional model for the turbulent
  tachocline}. {\itshape \solphys} 2003, \textbf{215}, 17--30.

\bibitem[\protect\citeauthoryear{{Petrovay}}{2007}]{Petrovay:bimodal}
{Petrovay}, K., {On the possibility of a bimodal solar dynamo}. {\itshape
  Astron.~Nachr.} 2007, \textbf{328}, 777--780.

\bibitem[\protect\citeauthoryear{{Petrovay} and
  {Kerekes}}{2004}]{Petrovay+Kerekes:intflow}
{Petrovay}, K. and {Kerekes}, A., {The effect of a meridional flow on Parker's
  interface dynamo}. {\itshape Mon.~Not.~Roy.~Astr.~Soc.}
2004, \textbf{351}, L59--L62.

\bibitem[\protect\citeauthoryear{Sch{\"u}ssler}{1984}]{Schussler:vort.pump}
Sch{\"u}ssler, M., On the structure of magnetic fields in the solar convective
  zone; in {\itshape The Hydromagnetics of the Sun}, edited by T.D. Guyenne and
  J.J. Hunt 1984, pp. 67--76.

\bibitem[\protect\citeauthoryear{Solanki
  {\itshape{et~al.}}}{2006}]{Solanki+:RPP}
Solanki, S., Inhester, B. and Sc{h\"us}sler, M., The solar magnetic field.
  {\itshape Rep.~Prog.~Phys.} 2006, \textbf{69}, 563--668.

\bibitem[\protect\citeauthoryear{{Tobias}}{1996}]{Tobias:nonlin.interface}
{Tobias}, S.M., {Diffusivity quenching as a mechanism for Parker's surface
  dynamo}. {\itshape Astrophys.~J.}
1996, \textbf{467}, 870--880.

\bibitem[\protect\citeauthoryear{Zeldovich
  {\itshape{et~al.}}}{1984}]{Zeldovich+:mg.book}
Zeldovich, Y.B., Ruzmaikin, A.A. and Sokoloff, D.D., {\itshape Magnetic Fields
  in Astrophysics},  1984 (New York: Gordon \& Breach).

\bibitem[\protect\citeauthoryear{{Zhang}
  {\itshape{et~al.}}}{2004}]{Zhang+:sandwich.dynamo}
{Zhang}, K., {Liao}, X. and {Schubert}, G., {A sandwich interface dynamo:
  Linear dynamo waves in the Sun}. {\itshape Astrophys.~J.}
2004, \textbf{602},
  468--480.

\end{thebibliography}

\end{document}